\documentclass{article}[12pt]
\usepackage{geometry,amsmath,amssymb,graphicx,calc,capt-of,ifthen}
\newcommand{\email}[1]{{\textit{Email:} \texttt{#1}}}
\newcommand{\tmmathbf}[1]{\ensuremath{\boldsymbol{#1}}}
\newcommand{\tmop}[1]{\ensuremath{\operatorname{#1}}}
\newcommand{\tmstrong}[1]{\textbf{#1}}
\newcommand{\tmtextit}[1]{{\itshape{#1}}}
\newcommand{\tmfloatcontents}{}
\newlength{\tmfloatwidth}
\newcommand{\tmfloat}[5]{
  \renewcommand{\tmfloatcontents}{#4}
  \setlength{\tmfloatwidth}{\widthof{\tmfloatcontents}+1in}
  \ifthenelse{\equal{#2}{small}}
    {\ifthenelse{\lengthtest{\tmfloatwidth > \linewidth}}
      {\setlength{\tmfloatwidth}{\linewidth}}{}}
    {\setlength{\tmfloatwidth}{\linewidth}}
  \begin{minipage}[#1]{\tmfloatwidth}
    \begin{center}
      \tmfloatcontents
      \captionof{#3}{#5}
    \end{center}
  \end{minipage}}

\begin{document}

\title{The Cummings-Stell model of associative fluids:  \ \ \ \ \ \ \ \ \ \ \
\ \ \ \ \ \ \ \ a general solution}\author{J. F.
Rojas\thanks{\email{frojas@fcfm.buap.mx; }}\\
}\maketitle

\begin{abstract}
  In a series of publications the Cummings-Stell model (CSM), for a binary mixture
  of associative fluids with steric effects, has been solved analytically
  using the Percus-Yevick approximation (PYA). The solution consists in a
  square well potential of width $w$, whose center is placed into the hard
  sphere shell ($r < \sigma$): at $L = \sigma / n$ ($n = 1, \ldots, 4$). This
  paper presents a general solution, for any $n$, of the first order
  Difference Differential Equation (DDE), for the auxiliary Baxter's function
  that appears in the CSM, using recursive properties of these auxiliary
  functions and a matrix composed by differential and shift operators (MDSO).
  This problem is common in some other models of associative fluids such as
  the CSM for homogeneus and inhomogeneus mixtures of sticky shielded hard
  spheres including solvent effects under PYA, and in that of mean-spherical
  approximation (MSA), for chemical ion association and dipolar dumbbells and
  polymers. The sticky potential implies a discontinuity step at $L$ in the
  solution of auxiliary Baxter's functions so that, one side, $L$ now is
  arbitrary and, for some additional effects, it can be placed one or more
  sticky potentials at different positions into the hard shell.
\end{abstract}


{\tableofcontents}

\section{Introduction}

In 1984 Cummings and Stell {\cite{aCS_1}} proposed a simplified hamiltonian
model for a reactive system of two types of homogeneus fluids with the same
density and diameter{\footnote{It will be detailed in next section.}}. In this
model, type $A$ and $B$ molecules, can be associated by means of a selective
square interaction that, in the appropiate limits, can be reduced to a sticky
Baxter's potential {\cite{abaxtersticky68}} located inside the hard core at a
distance from the center of the particle $L = \sigma / n$. The cases of $n =
2, 3, 4$ for the Baxter's function have been solved in
{\cite{aCS_1,aCS_2,aCS_3}} using the same formalism and, at {\cite{aCS_4}},
they apply the results for a pair of reactive fluids and a solvent. Lee and
Rasaiah solved the $L = \sigma / 4$, $L = \sigma / 5$ and proposed a solution
for $\sigma / n$ for chemical association and dipolar dumbbells
{\cite{aCSleerasaiah}}.

The model of Lee and Rasaiah of association in electrolytes $A^+ + B^-
\rightleftarrows A B$ is studied in {\cite{aleerasaiahelectro}}. In this case
the authors add a selective coulombian part, $\pm e^2 / r$, out of core, to
the associating original CSM.

The sticky site inside the hard core incorporates geometrical conditions of
steric saturation in the molecule and this idea is shown, using computational
simulations in different ensembles, in ref. {\cite{ahuerta}}. For different
bonding length parameters the system allows formation of dimers for small $L$,
chains for $L$ slightly larger, and vulcanization of species for bonding
length values close to the diameter $\sigma$ of particles. Huerta and Naumis
studied the connectivity of a binary mixture using a selectively hard sphere
potential and a superposition as {\cite{ahuertanaumis}}:
\[ U_{i j} (r) = U^{\tmop{hd}}_{i j} (r) + (1 - \delta_{i j}) U_{\tmop{as}}
   (r) \]
where
\[ U^{\tmop{hd}}_{i i} (r) = \left\{ \begin{array}{cc}
     \infty, & r < 1\\
     0, & r > 1
   \end{array} \right. \]
\[ U^{\tmop{hd}}_{i j} (r) = \left\{ \begin{array}{cc}
     \infty, & r < L - 0.5 w\\
     D, & L - 0.5 w < r < 1\\
     0, & r > 1
   \end{array} \right. \]
and
\[ U_{\tmop{as}} (r) = \left\{ \begin{array}{cc}
     0, & r < L - 0.5 w\\
     - \varepsilon_{\tmop{as}} - D, & L - 0.5 w < r < l + 0.5 w\\
     0, & r > L + 0.5 w,
   \end{array} \right. \]
where $L$ is the bonding distance, $w$ the intracore square well width and
$i$, $j$ represents the species in the mixture. The final potential (see
Figure \ref{potential}) is equivalent to the original Cummings-Stell in the
adecuate limits for sticky approximation.

The same idea is implemented by Pizio and Blum {\cite{apizioblum}} for a
hard-sphere fluid with dimerization $A + A \rightleftharpoons A_2$. In the
development most of the models maintain $L$ as a parameter (bonding distance)
and finally take the case $L = \sigma / 2$, however some other possibilities
are presents having the analytical solution for arbitrary $L$. Kalyuzhnyi and
Stell {\cite{ayustell}} present a recount of cases for different ranges of the
location $L$. As we show here, the sticky potential into the hard sphere shell
produces a discontinuity step in the auxiliary functions of Baxter. This fact
allows us to think of systems with more than one sticky site inside the shell
or, even, a distribution of sticky wells.

\begin{center}
  \tmfloat{h}{small}{figure}{\includegraphics{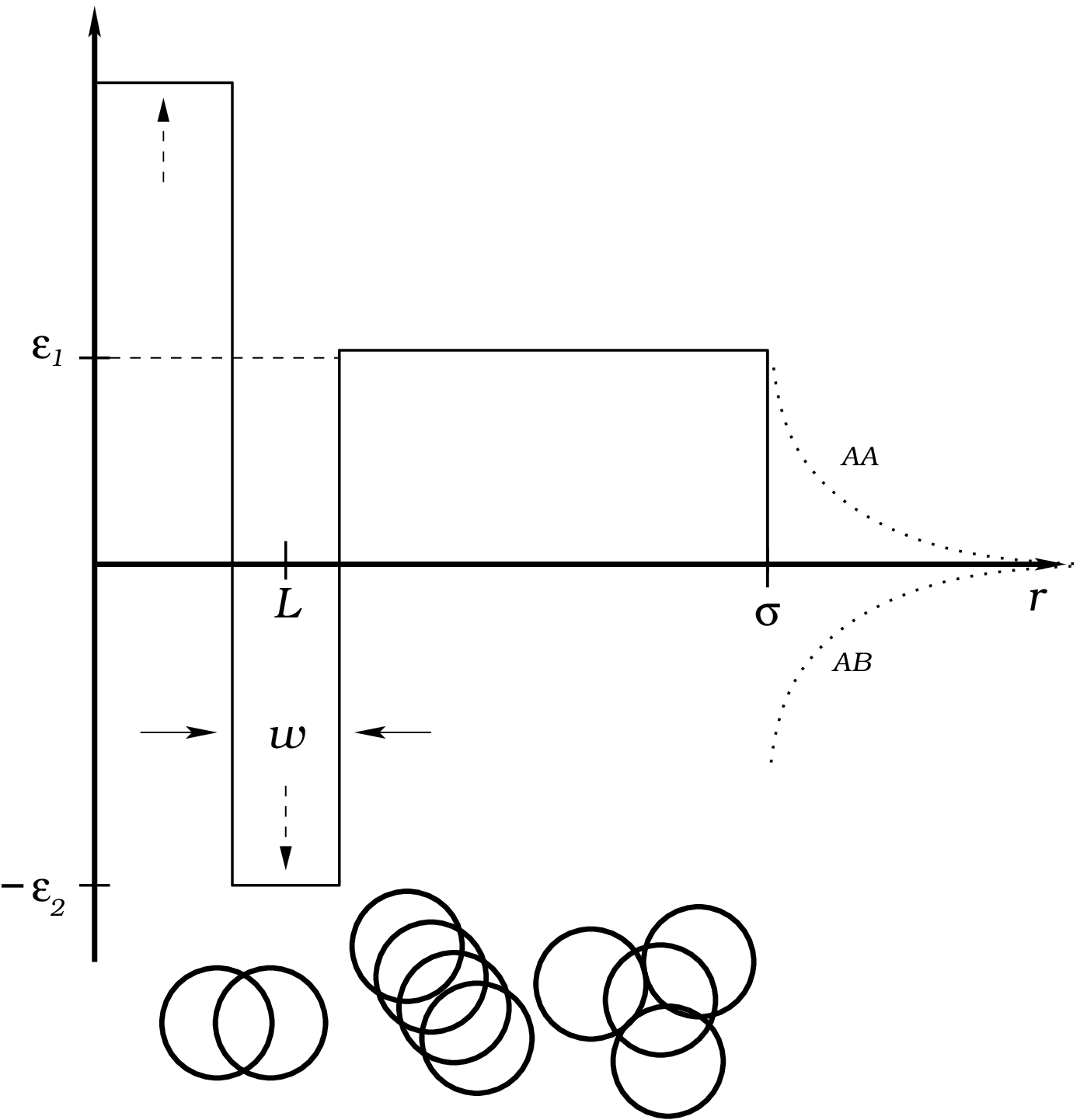}}{\label{potential}The
  image shows the basic form of the potential. Below,
  their structural effects for different $L$ according to
  {\cite{ahuertanaumis}}.}
\end{center}

The vertical dashed arrows show the sense of the Baxter's sticky limit
{\cite{aCS_1}}. The molecular diameter is $\sigma$ and the dashed curves (with
labels ``AA'' and ``AB'') corresponds to the colulombian interaction for
electrolyte {\cite{aleerasaiahelectro}}.

In the following sections, we first develop the matrix of differential and shift
operators MDSO, and its inverse, for the simplest case $n = 2$, followed for
the inversion
of the MDSO's for a very general case that corresponds to the set of $n$ DDE's, to $n$ subintervals of [0, $\sigma$] and to a sticky location
$L = m \sigma / n$.

\section{The association model: hard spheres with shielded sticky interaction}

\subsection{The model of binary mixture}

The statistical mechanical model of chemical reactions of Cummings and Stell
{\cite{aCS_1}} represents the association of two species $A + B
\rightleftarrows A B$, with the same density and diameter, which simplifies
the mathematical problem. The potential proposed in the CSM consists of a hard
sphere repulsion between like species ($A$-$A$ or $B$-$B$) and a mound of
width $\sigma$ with a deep, narrow, and attractive square well, with width $w$
centred on $L$. Here $L < \sigma / 2$ and $L + w / 2 \leqslant \sigma / 2$ for
$A B$ interactions:
\begin{equation}
  \phi_{A B} / k_B = \left\{ \begin{array}{ll}
    \epsilon_1 & \tmop{if} 0 < r < L - w / 2\\
    - \epsilon_2 & \tmop{if} L - w / 2 < r < L + w / 2\\
    \epsilon_1 & \tmop{if} L + w / 2 < r < \sigma\\
    0 & \tmop{if} r > \sigma
  \end{array} \right.
\end{equation}
The geometric consideration of this model for the $A B$ interactions ensures
steric saturation in the system (there is no formation of $n$-mers for $n
\geqslant 3$) due to overlapping. In addition, this model has a solution in
the PY approximation, mapping the square well onto an infinitely deep and
stretch well like the sticky potential of Baxter {\cite{abaxtersticky68}}. The
connection between the Baxter's original and this model, is obtained equating
the second virial coefficients: first it is considered the limit $_{}
\epsilon_1 \rightarrow \infty$, $\epsilon_2 \rightarrow \infty$, $w
\rightarrow 0$. The limits of $\epsilon_2 \rightarrow \infty$ and $w
\rightarrow 0$ are taken to maintain tractable the problem in the PY
approximation {\cite{abaxtersticky68}}. The limit $\epsilon_1 \rightarrow
\infty$ (in the repulsive part: see figure \ref{potential}) doesn't change
very much the results, but simplifies the solution {\cite{aCS_2}}.

The total and direct correlation functions are related by the Ornstein-Zernike
(OZ) equation that, for this binary mixture, can be written as
{\cite{aozernike}}
\begin{equation}
  h_{i j} (r) = c_{i j} (r) + \sum_{k = A, B} \rho_{k_{}, 0} \int c_{i k} (s)
  h_{_{k j}} (|\tmmathbf{r}-\tmmathbf{s}|) d\tmmathbf{s} \label{OZij}
\end{equation}
where the integral is evaluated in the whole space, $s =\|\tmmathbf{s}\|$, and
$\rho_{k, 0}$ is the number density of species $k$ particles. In the CSM the
densities are considered equal.

Briefly reviewing the formulation of the CS model: we need the factorized form
of the OZ {\cite{acorrfuncI}} equations (\ref{OZij}) with the PY closures
\begin{equation}
  \left. \begin{array}{ll}
    h_{A A} (r) = h_{B B} (r) = - 1, & r < \sigma\\
    & \\
    c_{A A} (r) = c_{B B} (r) = 0, & r > \sigma
  \end{array} \right\} \label{PY1}
\end{equation}
and, in the considered limits
\begin{equation}
  \left. \begin{array}{ll}
    h_{A B} (r) = - 1 + \frac{\lambda L}{12} \delta (r - L), & r < \sigma\\
    & \\
    c_{A B} (r) = 0, & r > \sigma .
  \end{array} \right\} \label{PY2}
\end{equation}
Given the conditions of the problem, the factorized OZ equations are written
{\cite{akinam,aCS_1,aCS_2,aCS_3}}
\begin{equation}
  \begin{array}{lll}
    r h_{11} (r) & = & - q_{11}' (r) + 2 \pi \rho \int^{\sigma}_0 d t (r - t)
    \left[ q_{11} (t) h_{11} (|r - t|) + q_{12} (t) h_{12} (|r - t|) \right]\\
    &  & \\
    r h_{12} (r) & = & - q_{12}' (r) + 2 \pi \rho \int^{\sigma}_0 d t (r - t)
    \left[ q_{11} (t) h_{12} (|r - t|) + q_{12} (t) h_{11} (|r - t|) \right] .
  \end{array} \label{OZfac}
\end{equation}
where we changed the index of $h_{A A}$ or $h_{A B}$ to $h_{11}$ or $h_{12}$
respectively. The same for all functions.

Substituting the closure relations (\ref{PY1}) and (\ref{PY2}) in the set of
OZ equations (\ref{OZfac}), the following system of difference-differential
equations (DDE) for the auxiliary Baxter's functions $q_{i j} (r)$ is
obtained:
\begin{equation}
  \begin{array}{lll}
    q_{11}' (r) + p [q_{12} (r + L) - q_{12} (r - L)] & = & (a_{11} + D
    a_{12}) r + b_{11} + D b_{12}\\
    &  & \label{qijs}\\
    q_{12}' (r) + p [q_{11} (r + L) - q_{11} (r - L)] & = & (D a_{11} +
    a_{12}) r + D b_{11} + b_{12} - \frac{\lambda L^2}{12} \delta (r - L),
  \end{array}
\end{equation}
where $p = \pi \rho \lambda L^2 / 6$ and
\[ \begin{array}{lll}
     a_{i j} & = & \delta_{i j} - 2 \pi \rho \int^{\sigma}_0 q_{i j} (t) d t\\
     &  & \\
     b_{i j} & = & 2 \pi \rho \int^{\sigma}_0 t q_{i j} (t) d t
   \end{array} \]
and satisfies the boundary conditions
\begin{equation}
  \begin{array}{ll}
    q_{11} (\sigma) = q_{12} (\sigma) = 0 & \\
    & \\
    q_{12} (L^-) = q_{12} (L^+) + \frac{\lambda L^2}{12} . & 
  \end{array} \label{condiciones}
\end{equation}
After integration, a step appears in $r = L$ for the auxiliary function
$q_{12} (r)$ due to the delta term associated with the well.

\subsection{Solving for $n = 2$}

{\noindent}In the first work of Cummings and Stell {\cite{aCS_1}} a new pair
of functions were used and defined as the sum and difference of the originals
$q_{11} (r)$ and $q_{12} (r)$. The advantage of this trick is to obtain two
uncoupled equations, one for $q_+ (r)$ and another for $q_- (r)$ which can be
solved in a separate way. If the functions $q_+ (r) = q_{11} (r) + q_{12} (r)$
and $q_- (r) = q_{11} (r) - q_{12} (r)$ are defined then, adding and
substracting Eqs. (\ref{qijs}),
\begin{equation}
  q_+' (r) + p [q_+ (r + L) - q_+ (r - L)] = a_+ r + b_+ - \frac{\lambda
  L^2}{12} \delta (r - L) \label{qmas}
\end{equation}
and
\begin{equation}
  q_-' (r) - p [q_- (r + L) - q_- (r - L)] = a_- r + b_- + \frac{\lambda
  L^2}{12} \delta (r - L), \label{qmenos}
\end{equation}
with the obvious definitions{\footnote{From {\cite{aCS_3}} $D = 1$, so that
$a_- = b_- = 0$. This not implies changes in the results. We asume this fact
in the rest of paper.}}
\[ a_{\pm} = (1 \pm D) \left[ 1 - 2 \pi \rho \int^{\sigma}_0 d t q_{\pm} (t)
   \right] \]
and
\[ b_{\pm} = (1 \pm D) 2 \pi \rho \int^{\sigma}_0 d t t q_{\pm} (t) . \]
The last term in (\ref{qmas}) and (\ref{qmenos}) can be omitted since it is
equal to zero in all subintervals except where $r = L = \sigma / n$. This
condition is fixed in the boundary conditions (\ref{condiciones}). In the rest
of subintervals, for the general case, must be true that
{\cite{aCS_1,aCSleerasaiah}}
\begin{equation}
  q (m \sigma / n^-) = q (m \sigma / n^+), \tmop{for} m = 2, 3, \ldots, n - 1.
  \label{diez}
\end{equation}

{\noindent}We use here $q (r) \equiv q_{11} (r) + q_{12} (r)$. The aim of this
proposal is to find an analytical form for the function $q (r)$ assuming that:
i) the solution must be made in subintervals {\cite{bbellmancooke}}, ii) this
implies that $q (r)$ will be defined also in subintervals, and iii) the
original functions $q_{i j} (r)$ can be recovered: $q_{11} (r) = \frac{q_+ (r)
+ q_- (r)}{2}$ and $q_{12} (r) = \frac{q_+ (r) - q_- (r)}{2}$. Identical
procedure shows that is sufficient to replace $\lambda \rightarrow - \lambda,
p \rightarrow - p, \nu \rightarrow - \nu$ to obtain $q_- (r)$.

\section{The MDSO}

The cases shown in {\cite{aCS_1,aCS_2,aCS_3,aCSleerasaiah}} are solved here
using MDSO. For convenience we show the case $L = \sigma / 2$ of the CS model
{\cite{aCS_1}} in detail, and the cases $L = \sigma / 3$ and $L = \sigma / 4$
summarized.

\subsection{The case $n = 2$}

The first case yields the system of coupled differential equations
\begin{equation}
  \frac{d q_1 (r)}{d r} + p q_2 (r + \sigma / 2) = a r + b, \tmop{for} 0 < r <
  \sigma / 2 \label{mdsoa1}
\end{equation}
and
\begin{equation}
  \frac{d q_2 (r)}{d r} - p q_1 (r - \sigma / 2) = a r + b, \tmop{for} \sigma
  / 2 < r < \sigma . \label{mdsob1}
\end{equation}
where, evidently, $q_1$ corresponds to the first half of the interval and
$q_2$ to the second. We define here the differential operator $\mathcal{D}$ as
$\mathcal{D}f (x) \equiv \frac{d f (x)}{d x}$ and the shift operator
$\mathcal{E}^s$ by $\mathcal{E}^{\pm s} f (x) \equiv f (x \pm s)$. With this
operators defined, the set of (\ref{mdsoa1}) and (\ref{mdsob1}) can be
rewritten as
\[ \mathcal{D}q_1 (r) + p\mathcal{E}^{\sigma / 2} q_2 (r) = a r + b \]
and
\[ \mathcal{D}q_2 (r) - p\mathcal{E}^{- \sigma / 2} q_1 (r) = a r + b \]
or, in matricial form, as
\begin{equation}
  \left( \begin{array}{cc}
    \mathcal{D} & p\mathcal{E}^{\sigma / 2}\\
    - p\mathcal{E}^{- \sigma / 2} & \mathcal{D}
  \end{array} \right) \left( \begin{array}{c}
    q_1 (r)\\
    q_2 (r)
  \end{array} \right) = \left( \begin{array}{c}
    f_1 (r)\\
    f_2 (r)
  \end{array} \right) . \label{mdsocero}
\end{equation}
These equations can be reduced to a symbolic form
\begin{equation}
  \mathcal{M}_2 \tmmathbf{q}(r) =\tmmathbf{f}(r) \label{formalset}
\end{equation}
where $\mathcal{M}_2$ is the matrix of differential and shift operators, or
MDSO, that appears in (\ref{mdsocero}), applied to the vector $\tmmathbf{q}$
of functions $q_i (r)$. The right side is the vector $\tmmathbf{f}$ of
functions $f_i (r)$ that, in this case, are linear functions of $r$. The index
in $\mathcal{M}$, corresponds to the number of equations (or partitions in the
interval of solution).

{\noindent}The main idea of this paper is to find a solution for the system
represented in (\ref{formalset}) as
\[ \tmmathbf{q}(r) =\mathcal{M}_2^{- 1} \tmmathbf{f}(r) . \]
This implies the knowledge of an explicit analytical form of the inverse of
$\mathcal{M}_2$, and how it operates on $\tmmathbf{f}(r)$. One way of defining
the inverse of the differential operator $\mathcal{D}$ is by using the
equation
\begin{equation}
  y' (x) \pm a y (x) = f (x) \label{ecdif}
\end{equation}
or
\[ (\mathcal{D} \pm a) y (x) = f (x) \]
whose solution leads us to define the inverse operator $(\mathcal{D} \pm a)^{-
1}$ as
\begin{equation}
  (\mathcal{D} \pm a)^{- 1} f (x) \equiv C e^{\mp a x} + e^{\mp a x}  \int
  e^{\pm a x'} f (x') d x' . \label{defineop}
\end{equation}
In the previous expression, the case $a = 0$ implies that the inverse MDSO is
reduced to the trivial definition of inverse differential operator as an
integral operator. The case where $a$ is a complex number (or a pure imaginary
one) implies harmonic solutions {\cite{pmosaicosmat}} and Fourier transform of
the right hand side of differential equation.

{\noindent}Continuing with the case $L = \sigma / 2$, the inverse of
$\mathcal{M}_2$ is
\begin{equation}
  \mathcal{M}^{- 1}_2 = \frac{1}{\Delta_2}  \left( \begin{array}{cc}
    \mathcal{D} & - p\mathcal{E}^{- \sigma / 2}\\
    p\mathcal{E}^{\sigma / 2} & \mathcal{D}
  \end{array} \right) \equiv \frac{1}{\Delta_2}  \check{\mathcal{M}}_2 
  \label{m2inverse}
\end{equation}
where the commutation properties of the operators $\mathcal{D}$ and
$\mathcal{E}$ were used. Direct calculation gives the determinant-operator of
$\mathcal{M}^{- 1}_2$ as{\footnote{The application of inverse shifting
operators is the identity: $\mathcal{E}^s \mathcal{E}^{- s} f (x)
=\mathcal{E}^s f (x + s) = f (x + s - s) = f (x)$.}}
\begin{equation}
  \Delta^{}_2 \equiv \mathcal{D}^2 + p^2 = (\mathcal{D}+ i p)^{} (\mathcal{D}-
  i p)^{} \label{detinverse},
\end{equation}
so that (\ref{m2inverse}) and (\ref{detinverse}) define completely the inverse
determinant-operator of $\mathcal{M}_2$ as the product of two inverse
operators of the form of (\ref{defineop}):
\begin{equation}
  \frac{1}{\Delta_2} = \Delta^{- 1}_2 = \frac{1}{\mathcal{D}+ i p} 
  \frac{1}{\mathcal{D}- i p} . \label{deltainverse}
\end{equation}

{\noindent}This is the formal inverse determinant of the MDSO, however we
still need to find the appropiate coefficients to satisfy the boundary
conditions. So that, the direct application of the inverse MDSO,
(\ref{m2inverse}), on (\ref{mdsoa1}) and (\ref{mdsob1}) gives
{\cite{bbellmancooke,pmosaicosmat}}
\[ \begin{array}{lll}
     q_1 (r) & = & A \cos p r + B \sin p r - \frac{a}{p} r + \frac{a}{p^2} (1
     - \nu / 2) - \frac{b}{p}\\
     &  & \\
     q_2 (r) & = & C \cos p r + D \sin p r + \frac{a}{p} r + \frac{a}{p^2} (1
     - \nu / 2) + \frac{b}{p}
   \end{array} \]
with $\nu \equiv p \sigma$. Now, considering (\ref{condiciones}) and the fact
of (\ref{mdsob1}) must be satisfied we obtain{\footnote{This fact allows us to
establish the same set of constants for the harmonic part of the solution.}},
explicitly,
\begin{equation}
  \begin{array}{lll}
    q_1 (r) & = & A \cos p r + B \sin p r - \frac{a}{p} r + \frac{a}{p^2} (1 -
    \nu / 2) - \frac{b}{p}\\
    &  & \label{q1q2AB}\\
    q_2 (r) & = & A \sin p (r - \sigma / 2) + B \cos p (r - \sigma / 2) +
    \frac{a}{p} r + \frac{a}{p^2} (1 - \nu / 2) + \frac{b}{p}
  \end{array}
\end{equation}
which agree exactly with the results in {\cite{aCS_1}}. The second equation,
now has the same set of constants that the first. The harmonic functions have
been interchanged and their arguments are shifted by $- \sigma / 2$.

\subsection{The more general case $n > 2$}

In the case $L = \sigma / 3$ there are three equations{\footnote{The first and
last equations always have one term less, due to the condition of PYA, $q (r)
= 0$ out of $[0, \sigma]$.}}:
\[ \frac{d q_1 (r)}{d r} + p q_2 (r + \sigma / 3) = a r + b, \tmop{for} 0 < r
   < \sigma / 3 \]
\begin{equation}
  \frac{d q_2 (r)}{d r} + p q_3 (r + \sigma / 3) - p q_1 (r - \sigma / 3) = a
  r + b, \tmop{for} \sigma / 3 < r < 2 \sigma / 3 \label{case3}
\end{equation}
\[ \frac{d q_3 (r)}{d r} - p q_2 (r - \sigma / 3) = a r + b, \tmop{for} 2
   \sigma / 3 < r < \sigma \]
with the MDSO given as
\[ \mathcal{M}_3 = \left( \begin{array}{ccc}
     \mathcal{D} & p\mathcal{E}^{\sigma / 3} & 0\\
     - p\mathcal{E}^{- \sigma / 3} & \mathcal{D} & p\mathcal{E}^{\sigma / 3}\\
     0 & - p\mathcal{E}^{- \sigma / 3} & \mathcal{D}
   \end{array} \right) \]
and the inverse operator of $\mathcal{M}_3$
\[ \mathcal{M}^{- 1}_3 = \frac{1}{\mathcal{D}(\mathcal{D}^2 + 2 p^2)} \left(
   \begin{array}{ccc}
     \mathcal{D}^2 + p^2 & - p\mathcal{D}\mathcal{E}^{\sigma / 3} & p^2
     \mathcal{E}^{2 \sigma / 3}\\
     p\mathcal{D}\mathcal{E}^{- \sigma / 3} & \mathcal{D}^2 & -
     p\mathcal{D}\mathcal{E}^{\sigma / 3}\\
     p^2 \mathcal{E}^{- 2 \sigma / 3} & p\mathcal{D}\mathcal{E}^{- \sigma / 3}
     & \mathcal{D}^2 + p^2
   \end{array} \right) \]
where
\[ \frac{1}{\Delta_3} = \frac{1}{\mathcal{D}(\mathcal{D}^2 + 2 p^2)} =
   \frac{1}{(\mathcal{D}- 0)}  \frac{1}{(\mathcal{D}+ i \sqrt{2} p)} 
   \frac{1}{(\mathcal{D}- i \sqrt{2} p)} \]
and{\footnote{With this we are defining $\mathcal{M}^{- 1}_n =
\frac{1}{\Delta_n}  \check{\mathcal{M}_n}$.}}
\begin{equation}
  \check{\mathcal{M}}_3 = \left( \begin{array}{ccc}
    \mathcal{D}^2 + p^2 & - p\mathcal{D}\mathcal{E}^{\sigma / 3} & p^2
    \mathcal{E}^{2 \sigma / 3}\\
    p\mathcal{D}\mathcal{E}^{- \sigma / 3} & \mathcal{D}^2 & -
    p\mathcal{D}\mathcal{E}^{\sigma / 3}\\
    p^2 \mathcal{E}^{- 2 \sigma / 3} & p\mathcal{D}\mathcal{E}^{- \sigma / 3}
    & \mathcal{D}^2 + p^2
  \end{array} \right) \label{mgorro} .
\end{equation}

{\noindent}Obviously $1 / \Delta_3$ is a product of inverse operators in the
form of (\ref{defineop}). Applying these to the right hand side of
(\ref{case3}) we obtain, directly
\[ q_1 (r) = A_1 \cos \sqrt{2} p r + B_1 \sin \sqrt{2} p r + \frac{a}{2} r^2 -
   \frac{a}{2 p} (1 - 2 \nu / 3) r + b r + F_1 \]
\[ q_2 (r) = A_2 \cos \sqrt{2} p r + B_2 \sin \sqrt{2} p r + F_2 \]
\[ q_3 (r) = A_3 \cos \sqrt{2} p r + B_3 \sin \sqrt{2} p r + \frac{a}{2} r^2 +
   \frac{a}{2 p} (1 - 2 \nu / 3) r + b r + F_3 \]
and, imposing bound conditions in the respective subintervals and the fact of
(\ref{case3}) must be satisfied (as in (\ref{q1q2AB}) for $n = 2$), we obtain
$A_2, A_3, B_2$ and $B_3$ in terms of $A_1$ and $B_1$
\[ q_1 (r) = A_1 \cos \sqrt{2} p r + B_1 \sin \sqrt{2} p r + \frac{a}{2} r^2 -
   \frac{a}{2 p} (1 - 2 \nu / 3) r + b r \]
\begin{equation}
  q_2 (r) = \sqrt{2} A_1 \sin \sqrt{2} p (r - \sigma / 3) + \sqrt{2} B_1 \cos
  \sqrt{2} p (r - \sigma / 3) \label{sol3}
\end{equation}
\[ q_3 (r) = - A_1 \cos \sqrt{2} p (r - 2 \sigma / 3) - B_1 \sin \sqrt{2} p (r
   - 2 \sigma / 3) + \frac{a}{2} r^2 + \frac{a}{2 p} (1 - 2 \nu / 3) r + b r.
\]
The case $L = \sigma / 4$ has a tridiagonal matrix $\mathcal{M}_4$, whose
determinant $\Delta_4 =\mathcal{D}^4 + 3\mathcal{D}^2 p^2 + p^4$ has roots
$\pm i \frac{\sqrt{5} - 1}{2} p$ \ and $\pm i \frac{\sqrt{5} + 1}{2} p$ \ so
that
\[ \Delta_4 = \left[ \mathcal{D}+ i \frac{\sqrt{5} - 1}{2} p \right] \left[
   \mathcal{D}- i \frac{\sqrt{5} - 1}{2} p \right] \left[ \mathcal{D}+ i
   \frac{\sqrt{5} + 1}{2} p \right] \left[ \mathcal{D}- i \frac{\sqrt{5} +
   1}{2} p \right] . \]
Lee and Rasaiah, in {\cite{aCSleerasaiah}}, called these roots $x =
\frac{\sqrt{5} - 1}{2} $ and $y = \frac{\sqrt{5} + 1}{2}$, and correspond to
the $\alpha_k$'s defined below in this paper.

{\noindent}The case $n = 5$ or $L = \sigma / 5$ has a determinant $\Delta_5
=\mathcal{D}^5 + 4\mathcal{D}^3 p^2 + 3\mathcal{D}p^4$ whose roots are $0, \pm
i p, \pm i \sqrt{3} p$, and the solutions have the same structure as
(\ref{q1q2AB}) and (\ref{sol3}), and that showed in
{\cite{aCS_1,aCS_2,aCS_3,aCS_4}}.

\section{The general MDSO}

\subsection{The matrix}

Making the same construction for $n$ divisions in the solution interval, one
obtains $n$ functions $q_i$ that represent a continuous solution. Each of them
must be such that
\[ q_i' (r) + p [q_{i + 1} (r + \sigma / n) - q_{i - 1} (r - \sigma / n)] = a
   r + b, \tmop{for} \frac{i - 1}{n} \sigma < r < \frac{i}{n} \sigma \]
where $i = 1, 2, 3 \ldots, n$ and $q_0 = q_{n + 1} = 0$, in which
\begin{equation}
  q_i (r) = \left\{ \begin{array}{cc}
    q (r), & \frac{i - 1}{n} \sigma < r < \frac{i}{n} \sigma\\
    0, & \tmop{otherwise} .
  \end{array} \label{qider} \right.
\end{equation}
The generalized MDSO for arbitrary $n$ has the form tridiagonal
\begin{equation}
  (\mathcal{M}_n)_{i j} = \left\{ \begin{array}{c}
    \mathcal{D}\\
    p\mathcal{E}^{\sigma / n}\\
    - p\mathcal{E}^{- \sigma / n}\\
    0
  \end{array} \begin{array}{l}
    \tmop{for} i = j\\
    \tmop{for} i = j - 1\\
    \tmop{for} i = j + 1\\
    \tmop{oherwise}
  \end{array} \right.
\end{equation}
or, in matrix form
\begin{equation}
  \mathcal{M}_n = \left( \begin{array}{cccccc}
    \mathcal{D} & p\mathcal{E}^s & 0 & 0 & \ldots & 0\\
    - p\mathcal{E}^{- s} & \mathcal{D} & p\mathcal{E}^s & 0 & \ldots & 0\\
    0 & - p\mathcal{E}^{- s} & \mathcal{D} & p\mathcal{E}^s & \ldots & 0\\
    0 & 0 & \ldots & \ldots & p\mathcal{E}^s & 0\\
    \ldots & \ldots & \ldots & - p\mathcal{E}^{- s} & \mathcal{D} &
    p\mathcal{E}^s\\
    0 & 0 & 0 & 0 & - p\mathcal{E}^{- s} & \mathcal{D}
  \end{array} \right) \label{thematrix}
\end{equation}
Due to symmetry of $\mathcal{M}_n$ the shift operators are mutually canceled
in the inverse of $\Delta_n$. This fact enables us to put all solutions $q_i
(r)$ in terms of inverse differential operators of the form (\ref{defineop})
and (\ref{deltainverse}).

\subsection{The determinant inverse operator}

From the tridiagonal matrix obtained, (\ref{thematrix}), the determinants for
different values of $n$ can be evaluated:
\begin{eqnarray*}
  \Delta_0 & \equiv & 1\\
  \Delta_1 & = & \mathcal{D}\\
  \Delta_2 & = & \mathcal{D}^2 + p^2\\
  \Delta_3 & = & \mathcal{D}^3 + 2\mathcal{D}p^2\\
  \Delta_4 & = & \mathcal{D}^4 + 3\mathcal{D}^2 p^2 + p^4\\
  \Delta_5 & = & \mathcal{D}^5 + 4\mathcal{D}^3 p^2 + 3\mathcal{D}p^4\\
  \Delta_6 & = & \mathcal{D}^6 + 5\mathcal{D}^4 p^2 + 6\mathcal{D}^2 p^4 +
  p^6\\
  \ldots &  & \\
  \Delta_{14} & = & \mathcal{D}^{14} + 13\mathcal{D}^{12} p^2 +
  66\mathcal{D}^{10} p^4 + 165\mathcal{D}^8 p^6 + 210\mathcal{D}^6 p^8 +
  126\mathcal{D}^4 p^{10} +\\
  &  & + 28\mathcal{D}^2 p^{12} + p^{14}\\
  \Delta_{15} & = & \mathcal{D}^{15} + 14\mathcal{D}^{13} p^2 +
  78\mathcal{D}^{11} p^4 + 220\mathcal{D}^9 p^6 + 330\mathcal{D}^7 p^8 +
  252\mathcal{D}^5 p^{10} +\\
  &  & + 84\mathcal{D}^3 p^{12} + 8\mathcal{D}p^{14}\\
  \ldots &  & 
\end{eqnarray*}
{\noindent}It is easy to prove that the recurrence relation between
determinants of MDSO's of different order is, for $n \geqslant 2$,
\begin{equation}
  \Delta_n =\mathcal{D} \Delta_{n - 1} + p^2 \Delta_{n - 2}
\end{equation}
where $\Delta_1 \equiv \mathcal{D}$ and $\Delta_0 \equiv 1$. The index of
$\Delta_n$ corresponds to the order of the MDSO. From this recurrence relation
we obtain the general expression for $\Delta_n$
\[ \begin{array}{lll}
     \Delta_n & = & \mathcal{D}^n + \sum_{j = 1}^m  \frac{\mathcal{D}^{n - 2
     j} p^{2 j}}{j!}  \prod_{k = j}^{2 j - 1} (n - k)\\
     &  & \\
     & = & \sum_{j = 0}^m  \frac{\mathcal{D}^{n - 2 j} p^{2 j}}{j!}  \frac{(n
     - j) !}{(n - 2 j) !}
   \end{array} \]
which, finally, can be reduced to
\begin{equation}
  \Delta_n = \sum_{j = 0}^m \left( \begin{array}{c}
    n - j\\
    j
  \end{array} \right) \mathcal{D}^{n - 2 j} p^{2 j}
\end{equation}
where, additionally
\begin{equation}
  \Delta_n = \prod_{k = 1}^n (\mathcal{D}- x_k) . \label{deltafac}
\end{equation}
In these expressions $\left. \left. m = \right\lfloor n / 2 \right\rfloor$ is
the integral part of $n / 2$. In (\ref{deltafac}) we have written $\Delta_n$
in factorized polynomial form. Here $x_k$ are all the $n$ roots of $\Delta_n$
which are all pure imaginary and proportional to $p$. When $n$ is even there
are exactly $m = n / 2$ pairs of complex conjugated roots{\footnote{This was
proved for $n = 1, 2, \ldots, 26$.}} $\pm i \alpha_k p$ and, if $n$ is odd,
there is a further null root of $\Delta_n$, which requires an additional
integration to obtain the $q_i$ functions.

We can write $\Delta_n$, with $m$ as defined above, as
\begin{equation}
  \Delta_n = \left\{ \begin{array}{cc}
    \prod_{k = 1}^m \mathcal{D}^2_k & \tmop{for} n \tmop{even}\\
    & \\
    \mathcal{D}_0 \prod_{k = 1}^m \mathcal{D}^2_k & \tmop{for} n \tmop{odd}
  \end{array} \right.
\end{equation}
where we define $\mathcal{D}^2_k \equiv (\mathcal{D}- x_k) (\mathcal{D}-
\tilde{x}_k)$ and $x_k \equiv i s_k \equiv i \alpha_k p$. Also, $\tilde{x}_k$
is the complex conjugate of $x_k$, $\alpha_k$ is a real number, and
$\mathcal{D}_0$ is the operator associated with the root $x_0 = 0$ for $n$
odd.

Direct application of individual inverse $\mathcal{D}^2_k$ on a linear
function gives {\cite{pmosaicosmat}}
\begin{equation}
  \mathcal{D}^{- 2}_k (c r + d) = A_k \cos s_k r + B_k \sin s_k r +
  \frac{1}{s^2_k} (c r + d) .
\end{equation}
Same as before, $s_k$ is defined by mean of $x_k \equiv i s_k \equiv i
\alpha_k p$ and $D^{- 2}_k \equiv (D^2_k)^{- 1}$. In the general case, a
partition $n$ even of the interval [$0, \sigma$] results in
\begin{equation}
  \mathcal{D}^{- 2}_1 \mathcal{D}^{- 2}_2 \ldots \mathcal{D}^{- 2}_m (c r + d)
  = \sum_{k = 1}^m A_k \cos s_k r + B_k \sin s_k r + \frac{1}{\prod^m_{k = 1}
  s^2_k} (c r + d) . \label{d2minverse}
\end{equation}
It is easy to see that if we define the vectorial funcion $\tmmathbf{v}(r)$,
using the definition in (\ref{m2inverse}) or (\ref{mgorro}), as the
application of $\check{\mathcal{M}}_n$ to the vector $\tmmathbf{f}(r)$,
\[ \tmmathbf{v}(r) = \check{\mathcal{M}}_n \tmmathbf{f}(r), \]
we obtain a linear function $\tmmathbf{v}(r)${\footnote{The harmonic part
comes from the inverse determinant.}}. The general solution of the system of
equations become
\[ \tmmathbf{q}(r) = \frac{1}{\Delta_n} \tmmathbf{v}(r) \]
where
\[ \tmmathbf{q}(r) \equiv \left( \begin{array}{c}
     q_1 (r)\\
     q_2 (r)\\
     \ldots\\
     q_n (r)
   \end{array} \right), \]
and $q_i (r)$ is as defined in (\ref{qider}). The constants $A_i$ and $B_i$
that appears in (\ref{d2minverse}) can be stablished from bound conditions at
the frontiers of the subintervals given by (\ref{condiciones}) and
(\ref{diez}) and the recurrence stablished in the original set of equations.

\section{Discussion}

A general solution of the set of DDE's for the original CS model has been
discussed here. It requires the roots of the polinomial expresion for the
inverse determinant $\Delta^{- 1}_n$ which always are of the form $i s_k p$
with $s_k$ a real number. One of the advantages of this method is the fact
that it allows to choose the site of the sticky potential, not only at $L =
\sigma / n$ for a few values of $n$: the solution is valid for any $n$. It
allows to change the sticky potential site to positions more and
more close to (or away of) the center of the particle. For example in the case
$n = 3$ the step of discontinuity $\lambda L^2 / 12$ can be defined at $L =
\sigma / 3$ or at $L = 2 \sigma / 3$.

This fact bring us the possibility to set the well at $r = m \sigma / n$ to
obtain different molecular structures {\cite{ahuertanaumis}}: by locating the
step $\lambda L^2 / 12$ in the site of sticky and represent a step
discontinuity in the solution at the position $L = m \sigma / n$. With this in
mind one can think, also, the posibility of two or more sticky square wells
into the hard shell.\\

{\noindent}{\tmstrong{Acknowledgements}}

{\noindent}J. F. Rojas, is very grateful to Prof. L. Blum.

\end{document}